\documentclass[twocolumn,prl,showpacs,preprintnumbers,superscriptaddress,amsmath,amssymb]{revtex4} 
\usepackage{amsmath,dcolumn,graphicx,epsf,ulem}          
\setkeys{Gin}{width=8.5cm,keepaspectratio}          
\usepackage{dcolumn}
\usepackage{bm}

\begin{document}          
          
\title{The unusual magnetism of nanoparticle $\rm LaCoO_3$}    
   
\author{A.~M.~Durand}    
\affiliation{Department of Physics, University of California, Santa Cruz, CA 95064, USA}    
\author{D.~P.~Belanger}    
\affiliation{Department of Physics, University of California, Santa Cruz, CA 95064, USA}    
\author{T.~J.~Hamil}    
\affiliation{Department of Physics, University of California, Santa Cruz, CA 95064, USA}    
\author{F.~Ye}    
\affiliation{Quantum Condensed Matter Division, Oak Ridge National Laboratory, Oak Ridge, Tennessee 37831, USA}    
\author{S.~Chi}    
\affiliation{Quantum Condensed Matter Division, Oak Ridge National Laboratory, Oak Ridge, Tennessee 37831, USA}    
\author{J.~A.~Fernandez-Baca}    
\affiliation{Quantum Condensed Matter Division, Oak Ridge National Laboratory, Oak Ridge, Tennessee 37831, USA}    
\author{C.~H.~Booth}    
\affiliation{Chemical Sciences Division, Lawrence Berkeley National Laboratory, Berkeley, CA 94720, USA}    
\author{Y.~Abdollahian}  
\affiliation{Department of Chemistry, University of California, Santa Cruz, CA 95064, USA}  
\author{M.~Bhat}    
\affiliation{Castilleja School, Palo Alto, CA 94301, USA}    
    
\date{\today}    
    
\begin{abstract}    
    
Bulk and nanoparticle powders of $\rm LaCoO_3$ (LCO)    
were synthesized, and their magnetic and structural properties    
were studied using SQUID magnetometry and neutron diffraction.  
The bulk and large nanoparticles exhibit weak ferromagnetism  
(FM) below $T \approx 85$~K and a crossover from strong to  
weak antiferromagnetic (AFM) correlations near a transition  
expressed in the lattice parameters, $T_o \approx 40$~K. 
This crossover does not occur in the smallest nanoparticles;
instead, the magnetic behavior is predominantly ferromagnetic.
The amount of FM in the nanoparticles depends on the amount    
of $\rm Co_3O_4$ impurity phase, which induces tensile strain on
the LCO lattice. A core-interface model is introduced, with the  
core region exhibiting the AFM crossover and with FM in the  
interface region near surfaces and impurity phases.  
    
\end{abstract}    
    
\maketitle          
         
The unusual magnetic behavior of $\rm LaCoO_3$ (LCO) has remained   
largely unexplained, despite the growing realization   
that structural distortion represents an important   
degree of freedom influencing the behavior of a   
large class of perovskites\cite{srkkmsw12,gl11}.    
Recently, strain-switched ferromagnetism (FM) in LCO   
has been used to create a spintronic device.~\cite{hppjdy13}   
Although the temperature at which that device   
operates is low ($T<90$~K), understanding   
the mechanism behind strain-induced LCO magnetism should facilitate   
the search for similar perovskite materials that will 
allow  switching of the ferromagnetic moment 
at higher temperatures.  Finding such a material 
will allow construction of spintronic devices for 
widespread use.  Recently, a model for 
the magnetism was developed~\cite{Durand} that explains the 
crucial role that the $\rm Co_3O_4$ impurity phase plays 
in the formation of long-range ferromagnetic order 
in LCO.  The model involves two regions: the \textbf{interface} region 
near the boundaries between the LCO and $\rm Co_3O_4$ phases as well as 
near the LCO particle surfaces, 
and the \textbf{core} LCO region away from these interfaces and surfaces.  
In this work, we apply the model to explain the effects of the particle 
surfaces and $\rm Co_3O_4$ impurity phase on the LCO magnetism; these 
effects are more pronounced as the LCo particle size decreases 
to the nanoscale. 
   
Many earlier attempts to model LCO magnetism   
focused on local transitions   
between Co electron states.  Such models   
do not provide a comprehensive description   
of the variety of phenomena observed in films,   
bulk and nanoparticle powders, and single crystals   
of LCO.  More recent efforts recognize   
the importance of including collective behaviors   
of the correlated electrons.\cite{dbbycfb13,lh13,rkfk99}  
By considering four samples with different sized  
particles and $\rm Co_3O_4$ impurity phase concentrations, we  
show that the disparate magnetic behaviors   
observed in various LCO systems fit well into the 
model developed~\cite{Durand} for the bulk particles. 

\section*{Synthesis and Characterization}  
   
The LCO bulk sample, A, was synthesized   
using a standard solid state reaction \cite{lstrk99}. Stoichiometric amounts   
of $\rm La_2O_3$ and $\rm Co_3O_4$ were ground together thoroughly and fired for   
8 hours. This process was repeated five times, with firing temperatures   
between 850$^\circ$C to 1050$^\circ$C.   
   
LCO nanoparticles were synthesized using the amorphous       
heteronuclear complex DTPA.~\cite{ftyjyc00, jbsbamz09}   
A 1.0 M NaOH solution was added by drops to an aqueous solution of       
$\rm La(NO_3)_3 \cdot 6H_2O$ and $\rm Co(NO_3)_3 \cdot 6H_2O$    
to prepare hydroxides.       
A stoichiometric amount of NaOH was used for sample D, and   
excess Na ions were removed via dialysis over   
24 hours. This resulted in significant $\rm Co_3O_4$   
phase, likely as some La ions were removed along with the Na.   
Only a 12.5$\%$ stoichiometric amount   
of NaOH was added to samples B and C,    
and no dialysis was undertaken, resulting in less $\rm Co_3O_4$.    
For all cases, equimolar amounts of DTPA were then added to the metal hydroxides,   
and the resulting complex precursor was stirred while heated   
to 80$^\circ$C. The resulting transparent solution was vaporized slowly at 80$^\circ$C   
until a dark purple resin-like gel formed, which was decomposed in air at 350$^\circ$C        
for 1.5 hours. The ash-like       
material was then calcined for 4 hours at 620$^\circ$C for samples   
C and D and 1000$^\circ$C for sample B.     

X-ray scattering data for characterization were taken using a Rigaku 
SmartLab powder diffractometer equipped with a copper x-ray tube 
($\lambda$ (Cu K$_\alpha$) = 1.54056 $\rm \AA$, 
tube energy 44 mA / 40 kV). The samples were analyzed with a 
scan rate of 3.0 $^\circ$/min with a step size of 0.02 $^\circ$.
Scans showed that all samples were 
predominantly LCO phase, and contained varying amounts of the 
$\rm Co_3O_4$ phase; the latter were further quantified in the neutron 
scattering measurements.
Samples C and D were also examined using small angle x-ray scattering (SAXS) on 
the same x-ray diffractometer. The nanoparticles 
were placed in between two layers of scotch tape, and transmission SAXS
was performed. Particle size and distribution data were analyzed 
using the NANO-solver software included with the diffractometer.

The particles in A and B were too large for size determinations using either   
neutron or x-ray scattering. We estimate an average size of 500 nm for A and   
between 100 and 400 nm for B based on TEM measurements on a bulk powder and a   
similar $\rm La_{1-x}Sr_xCoO_3$ nanoparticle powder with a calcination   
temperature of 1000$^\circ$C, respectively. 

Standard Scherrer analysis 
of x-ray diffraction data for sample D 
yielded average crystallite sizes of 22 nm. SAXS measurements on the 
same sample yielded an average particle size of 65 nm, 
suggesting the agglomeration of LCO nanoparticle crystallites during growth.    
Scherrer analysis for sample C gave an average crystallite size of 
18 nm, and the SAXS gave an average agglomerate size of 53 nm.    
   
\begin{figure}          
\includegraphics[height=5.3in, angle=0]{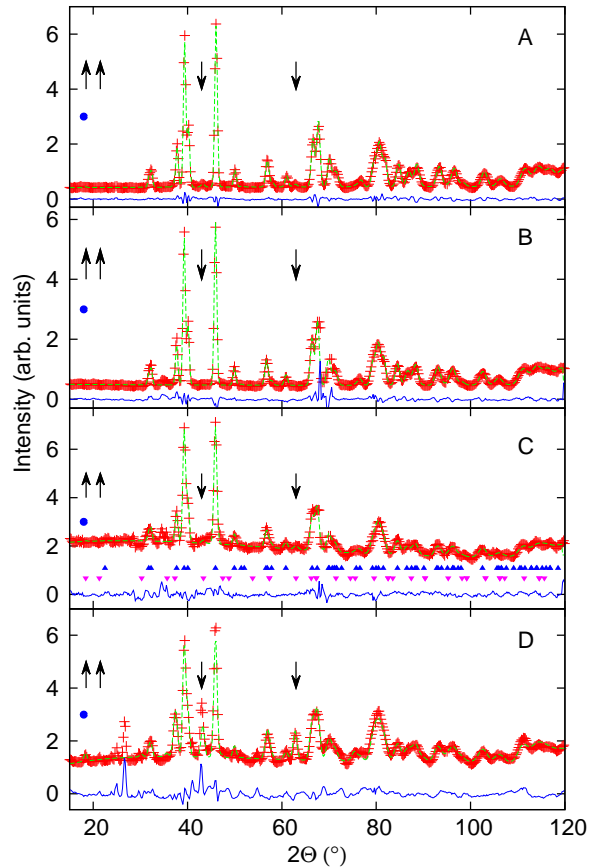}          
\caption          
{Neutron diffraction intensity vs $2\theta$ for $T=10$~K   
with FullProf refinements   
using the R$\overline{3}$c perovskite structure as well as the   
difference between the fit and calculation (lower curve)  
for samples A, B, C, and D.   
The upper row of triangles in panel C indicates calculated peak positions   
for LCO and the lower row of inverted triangles indicates peak positions for $\rm Co_3O_4$.   
The down arrows indicate observed $\rm Co_3O_4$ structural peaks and the   
up arrows indicate $\rm Co_3O_4$ magnetic peaks.  The filled dot indicates   
the magnetic peak position for $\rm CoO$.   
\label{fig:LCOnanobulk_refine}          
}          
\end{figure}         
   
\begin{table}   
\caption{Synthesis characterizations for samples A, B, C, and D.   
The methods are described in the text for the bulk (1), nanoparticle (2),   
and nanoparticle with dialysis (3). $T_{form}$ is the temperature at which   
the particles formed.  The determinations of the particle sizes and $\rm Co_3O_4$   
weight percentages are described in the text.   
}   
\begin{tabular}{l*{6}{c}}   
\hline 
&Method&$T_{form}(^\circ$C)&Size (nm)&&\%$\rm Co_3O_4$ \\
\hline   
A & 1 & 850-1050 & $\approx$500 && 4.5 \\   
B & 2 & 1000 & 100-400 && 0 \\   
C & 2 & 620  & 18 && 11 \\   
D & 3 & 620  & 22 && 28 \\   
\hline 
\end{tabular}   
\label{table:character}   
\end{table}   
  
\section*{Neutron Diffraction}

Figure~\ref{fig:LCOnanobulk_refine} shows the neutron   
powder diffraction intensity vs $2\theta$   
at $T = 10$~K from the WAND instrument at ORNL,   
along with FullProf refinements~\cite{r90} done 
using the R$\overline3$c symmetry for LCO. 
The distribution of small, non-spherical particles makes 
nanoparticle refinements more challenging. Bulk and nanoparticle powders 
exhibit small $\rm Co_3O_4$ and $\rm CoO$ phase peaks   
with $\rm F\overline{4}3m$ and $\rm Fm3m$ symmetries, respectively.    
The $\rm Co_3O_4$ structural and magnetic peaks were of high enough 
intensity to be refined by FullProf; the calculated weight percentages 
are shown in Table \ref{table:character}. The relatively sharp 
Bragg peaks seen for this phase indicate that the $\rm Co_3O_4$ 
forms crystallites, which are likely interspersed within the 
LCO nanoparticles and bulk. As only the $\rm CoO$ 
high intensity magnetic peak could be seen, refinements 
were not possible on this phase. It is therefore 
unlikely that there is a significant amount of $\rm CoO$ 
in these samples; in addition to the neutron scattering data, 
no indication is given in the magnetometry data of a 
significant $\rm CoO$ magnetic moment as was seen in the 
samples of Fita \textit{et al}.~\cite{fmmpwtvhvg08}

\begin{figure}   
\includegraphics[width=2.2in,angle=270]{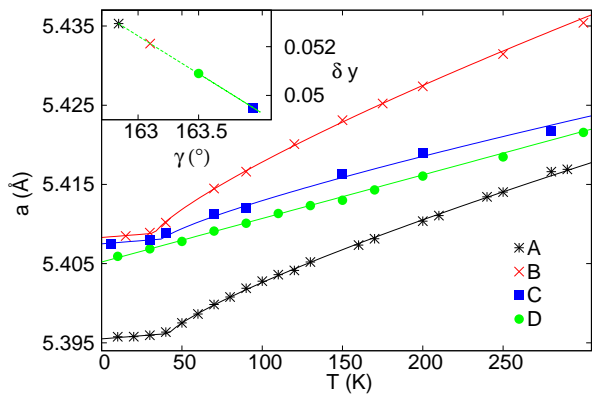}   
\caption   
{Lattice parameter $a(T)$ vs T for A, B, C and D.   
The inset shows $\delta y$ vs $\gamma$ for the same samples  
at $T=30$~K.   
\label{fig:LCO_a}   
}   
\end{figure}   
      
The bulk (A) average lattice parameter $a(T)$,    
shown in Fig.\ \ref{fig:LCO_a}, is significantly   
smaller than that of the nanoparticles (B, C and D).   
For $T>T_o$, A and B show significant    
curvature, whereas C shows only slight curvature and    
D shows no significant curvature.    
The inset of Fig.\ \ref{fig:LCO_a} shows the $T=30$~K  
values of $\gamma(T)$ (the Co-O-Co bond angle) and 
$\delta y(T)= \frac{d}{a}\cos (\gamma /2)$, where $d$ is the Co-O bond length. 
The parameter $\delta y(T)$ characterizes the rhombohedral distortion      
of the lattice~\cite{rkfk99,dbbycfb13,mkfoiyk99,lh13}.   
The parameters $\gamma$ and $\delta y$ are nearly proportional   
because $d$ varies by less than 0.4 \% across the samples at low $T$.   

The lattice parameter $a(T)$ for samples A, B and C, was fitted using   
      
\begin{equation}         
	a(T) = a_0(1 + \alpha T)  \quad \quad (T<T_o)   
\label{eq:linear}         
\end{equation}   
and   
\begin{equation}   
	a(T) = a(T_o) + K_a\bigg(\frac{T - T_o}{T_o}\bigg)^\sigma \quad \quad (T>T_o),   
\label{eq:power}   
\end{equation}   
         
{\noindent with the parameters in Table \ref{table:lattice_fit}. 
The temperature $T_o$ is the crossover point for the linear and power law 
behaviors, and has previously been suggested as the critical 
temperature for a collective phase transition in bulk LCO.~\cite{Durand}
Although more data are needed in the vicinity of the transition to ascertain 
the true value of $T_o$ in the B and C nanoparticle samples, we note that
the data are in qualitative agreement with the bulk LCO results. The data
shown in Fig.~\ref{fig:LCO_a} were fit with values of $T_o$ ranging from 
33 to 42 K. The sharp change in slope at $T_o$ is also observed in   
the lattice parameter $c(T)$ for A, B, and C. Sample D is better fit 
to Eq.\ \ref{eq:linear} over the entire $T$ range.   
   
\section*{Magnetometry}

\begin{table}
\caption{Fit parameters using Eq.\ \ref{eq:linear} and \ref{eq:power} with fixed exponent
$\sigma =0.8$. Errors in $\alpha$ and $K_a$ are similar for all samples.
} 
\begin{tabular}{l*{7}{c}}
\hline
& A && B && C && D \\
\hline
$a_0$(\AA)        &  5.396(1) && 5.409(1) && 5.406(1) && 5.407(1) \\
$T_o (K)$               & 42(2) && 33(2) && 38(2) && - \\
$\alpha$        &  3.0(5) $\times 10^{-6}$ && 3.0 $\times 10^{-6}$ && 3.0 $\times 10^{-6}$ && 5.5 $\times 10^{-5}$ \\
$K_a$         & 4.97(2) $\times 10^{-3}$ && 5.10 $\times 10^{-3}$ && 3.27 $\times 10^{-3}$ && - \\
\hline
\end{tabular}
\label{table:lattice_fit}
\end{table}

The behavior of the magnetization in LCO has long been considered    
unusual and yet no adequate model of it below room temperature    
has been developed.  Local Co spin-state models predict a nonmagnetic    
ground state because there are no moments when the spins 
are paired in the lowest energy state.  Moments are then thought to 
develop as spins are thermally excited near $T=90$~K. 
However, our data indicate that the ground state for material away from the
interfaces and particle surfaces 
clearly has magnetic moments at low temperature, though they do not order. 
Regions close to the interfaces or surfaces develop 
ferromagnetic order at low temperature. 
Recent work has shown the importance of extended   
states~\cite{lh13} in LCO bulk particles, a FM  
phase transition in LCO at $T_c \approx 87$~K, and   
another transition near $ T_o \approx 40$~K.~\cite{dbbycfb13} These behaviors,  
as well as others reported for bulk and nanoparticle powders and  
thin films,~\cite{yzg04_a,httski07,fapssl08,prkkmsfw09}   
are consistent with a particle core-interface model   
that includes, for particles larger than $\approx$ 20 nm, a core region exhibiting a   
crossover between two types of paramagnetism   
near $T_o$, and an interface region located near 
surfaces or interfaces with impurity phases. Tensile stress,   
either from the lattice mismatch between the 
interfaces and the core, or between the particle surface and the core, 
can induce a FM transition in the interface region below $T_c$.~\cite{fapssl08,fdaeaeksgl09}  

Figures \ref{fig:LCO_multi} and \ref{fig:LCO_multi_inv} show    
$M/H$ and $H/M$ vs $T$, respectively,    
for the $\rm LaCoO_3$ bulk and nanoparticle powders upon    
field cooling. Comprehensive fits using    
a simple superposition of FM and AFM    
behaviors were not successful. Instead, samples A, B and D   
were successfully fit using a superposition of two different Curie-Weiss-like (CW)   
paramagnetic behaviors and one power-law FM behavior,  
  
\begin{equation}     
\begin{split}    
\frac{M}{H}(T) =  \bigg(d + \frac{E_a}{T+t_a}\bigg)S(T) + \bigg(\frac{E_b}{T+t_b}\bigg) \\    
+ M_n\bigg(\frac{T_c-T}{T_c}\bigg)^\beta(S(T) + L(1-S(T))) ,    
\end{split}    
\label{eq:abd_fit}    
\end{equation} 
 
\noindent {for $T \le T_C$ and 
 
\begin{equation}     
\frac{M}{H}(T) =  \bigg(d + \frac{E_a}{T+t_a}\bigg)S(T) + \bigg(\frac{E_b}{T+t_b}\bigg)  
\label{eq:abd_fit_high}    
\end{equation} 
 
\noindent{for $T \ge T_C$, 
where the subscript $n$ indicates the field in Oe. Each term is modified by a sigmoid,  
centered at temperature $T_S$ with a width $1/W$,   
 
\begin{equation}     
S(T) = \frac{1}{1 + \exp(W(T_S-T))} .    
\label{eq:sigmoid}    
\end{equation}   
   
\noindent{This expression for $\frac{M}{H}(T)$ captures a crossover in the paramagnetic 
behavior at high $T$ (with parameters $E_a$ and $t_a$) 
to a low $T$  behavior ($E_b$ and $t_b$) as $T$ decreases.  
While the Curie-Weiss fits work well over the limited $T$ range used, the   
$E$ and $t$ parameters differ from high $T$ fits over the wide range $170<T<300$~K  
for the bulk.~\cite{dbbycfb13} The parameters cannot be directly interpreted 
as yielding the moment and interaction strength because the usual interpretation 
of the Curie-Weiss expression presumes weakly interacting moments. 
For sample C, which shows no evidence for a crossover or sharp 
transition to ferromagnetic order, but significant  
variation above $T_c$, data were fit to   
\begin{equation}     
	\frac{M}{H}(T) =  d_n + \frac{E_a}{T+t_a} + M_n\bigg(\frac{T_c-T}{T_c}\bigg)^\beta ,    
\label{eq:c_fit}    
\end{equation}    
\noindent{where the exponent for the power law is $\beta = 1.5$.  For this sample, the 
	parameter $d$ varied with the field.  For the fit to the data 
	for sample D, $t_a$ and $t_b$ were set equal. 
    
\begin{figure}
\includegraphics[width=3.1in,angle=0]{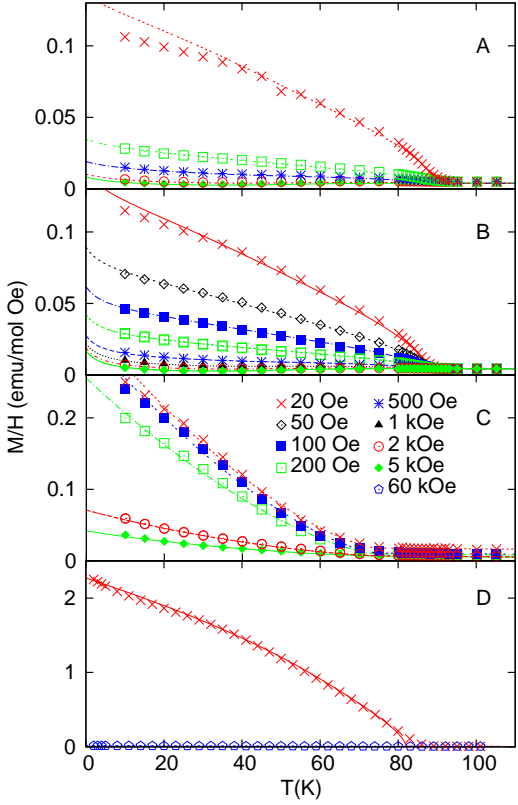}
\caption
{$M/H$ vs.\ $T$ for samples A, B, C and D for
fields $20<H<60000$~Oe along with fits described in
the text.
Note the different vertical scales for each data set.
\label{fig:LCO_multi}
} 
\end{figure}

\begin{figure}
\includegraphics[width=3.1in,angle=0]{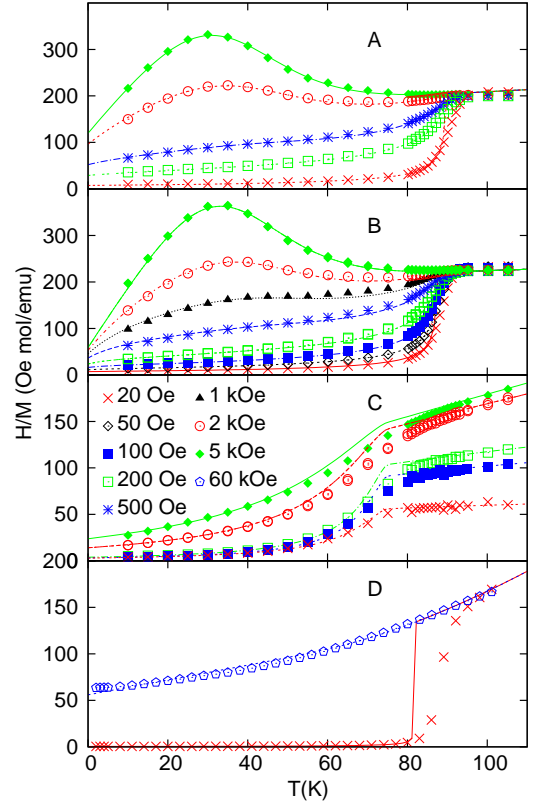}
\caption
{$H/M$ vs.\ $T$ for the same data and fits shown in Fig.\ \ref{fig:LCO_multi}.
\label{fig:LCO_multi_inv}
} 
\end{figure}

\begin{table}    
	\caption{Fit parameters for samples A, B, C, and D using   
          Eq.\ \ref{eq:abd_fit} and \ref{eq:c_fit} with fixed value $\beta=0.63$. 
          The parameter $d$ varies with $H$ only for sample C, for which values are listed in   
	  the last row.  The $M$ and $d$ subscripts refer to the applied field in Oe.     
	}    
	\begin{tabular}{l*{5}{l}}    
		\hline 
		&A&B&C&D \\    
		\hline    
		$T_C$(K) & 89.5(5) & 87.5(5) & 75(2) & 82(5) \\    
		$T_S$(K) & 49(2) & 47(2) & 50(10) & 80(20) \\    
		$W$(1/K) & 0.09(1) & 0.09 & 0.09 & 0.05 \\    
		$t_a$(K) &180(20)&70(5)&40(5)&80(20) \\    
		$t_b$(K) &11.5(5)& 5(1)&-&80(20) \\    
		$E_a$(erg K/mol Oe) &0.60(2)&0.15(5)&0.73(5)&0.88(5) \\    
		$E_b$(erg K/mol Oe) &0.094(1)&0.077(1)&-&1.5(1) \\    
		$L$ &1.2(2)&1.15(2)&1&1.4 \\    
		$d$(erg/mol Oe)&0.0019(2)&0.0036(1)&-&0.0004(1) \\    
		$M_{20}$(erg/mol Oe) &0.106(3)&0.11(1)&0.31(1)&1.61(2) \\    
		$M_{50}$ &-&0.063(2)&-&-\\    
		$M_{100}$&-&0.039(2)&0.29(1)&-\\    
		$M_{200}$&0.022(2)&0.022(2)&0.23(1)&-\\    
		$M_{500}$&0.0092(3)&0.0092(5)&-&-\\    
		$M_{1000}$&-&0.0043(1)&-&-\\    
		$M_{2000}$&0.0018(2)&0.0017(1)&0.052(1)&-\\    
		$M_{5000}$&0.0000(1)&0.00006(5)&0.023(1)&-\\    
		$M_{60000}$&-&-&-&0\\    
		\hline    
	\end{tabular}    
	\begin{tabular}{l*{9}{l}}    
		        $d_{20}$&&$d_{100}$&&$d_{200}$&&$d_{2000}$&&$d_{5000}$\\    
		0.0115(1)&&0.0046(1)&&0.0033(1)&&0.0007(1)&&0.00036(1) \\    
		\hline 
	\end{tabular}    
	\label{table:mag_fit}   
    
\end{table}    
   
Despite the large number of parameters, shown in Table \ref{table:mag_fit}  
at a given $H$, the fits are strongly   
constrained because only $M$ was allowed to   
vary with $H$ for samples A, B and D, and only $M$ and   
$d$ were allowed to vary for sample C.  The $H$-dependent parameters   
characterize the FM contributions. AFM contributions, associated   
with the CW functions, are expected to be insensitive   
to $H$ and dominate for large $H$. They are most apparent for $H/M$ vs $T$   
in Fig.\ \ref{fig:LCO_multi_inv}.  The FM  
power law is most dominant at small $H$ and is best seen   
in Fig.\ \ref{fig:LCO_multi}.  The FM fixed point is   
at $H=0$, so the power law behavior for $M/H$ vs $T$ is stronger and sharper   
as $H$ decreases.  

\section*{Discussion}
      
In several previous studies, FM in LCO nanoparticles has been      
attributed to FM ordering of the surface,~\cite{yzg04_a,httski07}      
surface-induced lattice strain,~\cite{fmmpwtvhvg08} and unit-cell    
expansion.~\cite{wzwf12,zhzgzs09}      
Yan \textit{et al.}~\cite{yzg04_a} found that the magnetic    
susceptibility of their      
bulk particles increased as the surface-to-volume ratio increased;    
as all of the samples were from the same single crystal sample, this    
indicates that surface effects increase FM in LCO.    
Harada \textit{et al.}~\cite{httski07} and Fita   
\textit{et al.}~\cite{fmmpwtvhvg08} found that a decrease in particle   
size correlates with an increase the net moment.   
These studies are consistent with a surface-induced    
tensile stress resulting from lattice expansion of the surface   
regions, and with tensile stress in thin films   
induced by substrates~\cite{fapssl08,lszzgwh13,pbsskzwld11,hrsd09}.     
   
To interpret the results of the fits to the magnetization and lattice   
parameters, we employ a core-interface model with two regions distinguished   
by the character of the magnetic interactions and the proximity to surfaces   
or interfaces.  The core region of the particles, far from any surface or 
interface, is particularly  
significant in the larger particles.  In this   
region, the dominant interaction is AFM. However, 
the peak seen near $T = 30$~K in the $H/M$ data for samples A and B 
is broad, not sharp as would be expected in the case of long-range 
AFM order. We thus conclude that the core region does not order 
antiferromagnetically. 

The AFM correlations 
vary with $\gamma$ (or, equivalently, $\delta y$),  
which crosses a critical value at $T_o \approx 40$~K. 
The high $T$ AFM correlations 
in the core region disappear below $T_o$, and the Curie-Weiss parameters 
$E_b$ and $t_b$ are smaller in this temperature range. 

While it might be reasonable to interpret this as a paramagnetic 
behavior with weaker moments and interactions, it was found~\cite{Durand} 
in the bulk particles that the extrapolation of the 
low $T$ Curie-Weiss expression to $T=0$ nearly coincides with 
the extrapolation to $T=0$ of the Curie-Weiss fit in the range 
$170<T<300$~K.  This could indicate that the moments and 
antiferromagnetic interactions remain the same and the 
correlations are short-ranged in both temperature regions. 
This, in turn, would imply that the system is highly 
frustrated for $T<T_o$. 
The core region has a large volume in samples A and B and is responsible for   
most of the antiferromagnetic contribution to $H/M$ for large $H$.   
However, the $T=0$ value of $H/M$ is significantly smaller in 
powder B.  This would be consistent with a smaller 
volume of the core region in B because moments are all near interfaces or 
surfaces. 
In the interface region, $\gamma$ never crosses the critical value as a 
result of tensile strain, and thus remains large enough to sustain 
long-range ferromagnetic order.~\cite{fapssl08}   
Interfaces with other phases, as well as the mismatch in lattice parameters  
of the core and interface regions, can be sources of tensile strain  
that can induce long-range ferromagnetic order.   
   
In the bulk particles of sample A, core region volumes are large   
and peaks in $H/M$ vs $T$ for large $H$ near $T_o$   
occur as the high $T$ antiferromagnetic paramagnetism   
gives way to low $T$ paramagnetism with weaker   
antiferromagnetic correlations.   
It was shown~\cite{Durand} that the standard procedure 
of growing bulk particles in ambient atmosphere 
results in unreacted $\rm Co_3O_4$ remaining in the sample.
Bulk sample A has about 4.5 \% $\rm Co_3O_4$ by weight, which 
could allow for LCO-$\rm Co_3O_4$ crystallite interfaces.  
In addition to the impurity phase interfaces, 
the weak ferromagnetic power law   
behavior is consistent with the relatively small volume   
of the interface region near particle surfaces. This power law   
behavior for all $H$ is followed quite well for $T_o<T<T_C$,   
but the ferromagnetic contribution to the   
magnetization at small $H$ falls significantly below the power law   
for $T<T_o$ (see Fig.~\ref{fig:LCO_multi}(A)).  
Notably, muon depolarization~\cite{gtpbwl06}   
peaks in the temperature region $T_o<T<T_C$, are   
consistent with a decrease in correlations below $T_o$. 

The exponent $\beta = 0.63$ describes   
the power law behavior near $T_C$ well, but is inconsistent with   
expected bulk ferromagnetic long-range ordering, for which   
$\beta << \frac {1}{2}$. However, Binder and Hohenberg (BH)~\cite{bh72,bh74}   
have predicted an exponent, $\beta=0.65$, for surface critical behavior    
which is consistent with our fits and, in turn,   
with the ferromagnetism being associated primarily with   
LCO surfaces and interfaces. Notably, BH surface magnetism   
is assisted by the bulk ordering and would normally   
be difficult to observe against the ordering moment of the bulk.   
For LCO, it is primarily the surface that orders ferromagnetically, making   
it possible to observe the weak 2D ferromagnetic ordering.   
Seo \textit{et al.}~\cite{spd12} suggest that the ferromagnetic moment near   
surfaces is a result of spin canting.  If so,   
moments near the surface must also order antiferromagnetically   
and the BH surface ordering mechanism would be associated   
with 2D antiferromagnetic ordering at the surface   
and 3D antiferromagnetism further away from the surface.   
The weakening of the antiferromagnetism in the core region 
and the concomitant weakening of the ferromagnetism 
below $T_o$ indicates they may be related, i.e. 
the core region plays a role in supporting the surface magnetism.  The weak ferromagnetic   
ordering in the bulk particles has yet to be observed using   
neutron scattering, so it is not surprising that weak   
antiferromagnetic ordering has also not yet been observed.   
LCO single crystals show behavior similar to the  
bulk particles, despite a small proportion of surface area~\cite{yzg04_a}.  
This may be due to significant twinning and defects  
that serve a similar role as free surfaces,  
resulting in similar submicron structures.  
   
The large nanoparticles of sample B show remarkably similar,   
but not identical, magnetic behavior to the bulk. This suggests   
that the relative proportions of the core and interface regions, are similar.  
Although the amount of $\rm Co_3O_4$ is less in the large nanoparticles, 
the smaller size of the particles results in more particle surface area, which can 
have a similar effect. 
   
Remarkable features of the small nanoparticles in   
sample C include the lack of a clear crossover of the   
antiferromagnetic behavior and a relatively weak   
structural signature of the transition near $T_o$.   
This suggests that core region moments occupy a   
relatively minor part of the sample; nearly   
all of the moments are close to the surface in the interface region.   
With the large surface area, one might expect the overall  
magnetic moment to be much larger than that of samples  
A and B.  However, without strain from a mismatch of  
core and interface region, ferromagnetic long-range order  
generated by the particle surfaces 
is greatly reduced in the interface region.  

Powder D differs primarily from C in the amount of $\rm Co_3O_4$, 
which is nearly three times larger in D. 
Apparently, the magnetic moment can be made much larger with  
the introduction of strain from $\rm Co_3O_4$. 
The particles in C and D are similar in size,   
but the introduction of $\rm Co_3O_4$   
results in even less core region volume than sample C.   
The lack of any apparent transition    
in the lattice parameters near $T_o$, the CW fits  
using the same temperature at large and small $T$, and  
the close tracking of the magnetization to  
the power law to low $T$ are consistent  
with insignificant core region volume.  
It has been shown that, in bulk   
particles,~\cite{Durand} the LCO/$\rm Co_3O_4$   
interfaces cause significant ferromagnetism,   
presumably because they introduce tensile strain into   
the LCO lattice. The particles in D are nearly all   
in the interface region and, with the introduction of tensile strain 
from $\rm Co_3O_4$,  
the ferromagnetic moment is an order of   
magnitude stronger than in any of the other   
powders.   
   
We have shown that the various magnetic behaviors of   
LCO powders are consistent with the presence of   
two kinds of magnetic regions, the interior  
core region of the larger particles  
and the interface region near surfaces or interfaces with other phases.  
The literature has examples of nanoparticle LCO,   
some showing phase transitions and others not.~\cite{zhzgzs09,wzwf12}  
In our model, these differences would reflect the density of   
interfaces with phases such as $\rm Co_3O_4$.   
This may explain, for example, the surprising 
results of Wei \textit{et al.}~\cite{wzwf12} which  
show a decrease in magnetization and $T_C$ with 
decreasing nanoparticle size. These are in contrast to the 
results by Fita \textit{et al.} and Yan \textit{et al.}, 
which show the opposite effect.~\cite{fmmpwtvhvg08, yzg04_a} 
   
This model will help in the interpretation of the   
magnetic behavior of large LCO crystals, which   
typically show large amounts of twinning, and LCO films   
grown on substrates.  It will be useful in comparing   
band structure simulations with magnetic and structural   
data.  Most importantly, it will aid the search for materials  
with switchable ferromagnetism that are 
suitable for making spintronic devices that operate 
above room temperature.  
  
We thank F. Bridges,   
A. Elvin, B. Harmon, J. Howe, A. P. Ramirez, and N. Sundaram for  
helpful discussions and$/$or assistance with measurements.  
 
The work at the High Flux Isotope Reactor at ORNL was supported by 
the DOE BES Office of Scientific User Facilities. 
Work at Lawrence Berkeley National Laboratory was 
supported by the Director, Office of Science (OS), 
Office of Basic Energy Sciences (OBES), of the U.S. Department of 
Energy (DOE) under Contract No. DE-AC02-05CH11231. 
Some X-ray data in this work were recorded on an instrument  
supported by the NSF Major Research Instrumentation (MRI)  
Program under Grant DMR-1126845.

\bibliography{magnetism_thesis.bib}          
          
\end{document}